\documentstyle[sprocl]{article}
\def\Journal#1#2#3#4{{#1} {\bf #2}, #3 (#4)}

\def\PLB{{\em Phys. Lett.}  B}

\def\PRD{{\em Phys. Rev.} D}

\def\mco{\multicolumn}

\def\ra{\rightarrow}

\def\ko{K^0}

\def\be{\begin{equation}}
\def\ee{\end{equation}}
\def\bea{\begin{eqnarray}}
\def\eea{\end{eqnarray}}

\begin{document}
\title{NEUTRINO MASS TEXTURES AND THE NATURE OF NEW PHYSICS
IMPLIED BY PRESENT NEUTRINO DATA\footnote{Invited talk presented at the
Neutrino'96 conference, Helsinki; (to appear in the proceedings)}}
\author{ R. N. MOHAPATRA }
\address{Department of Physics, University of Maryland, College Park\\ 
Maryland 20742}

\maketitle\abstracts{
If all the indications for neutrino oscillations observed in the 
solar, atmospheric neutrino data as well as in the LSND experiment
are borne out by the ongoing and future experiments, then they
severely constrain the neutrino mass texture. In particular, the
need for an extra ultra-light sterile neutrino species is hard to avoid.
Such an extra neutrino has profound implication not only for physics
beyond the standard model but even perhaps for physics beyond conventional
grand unification. We discuss a scenario involving
a parallel (or shadow) universe that interacts with the familiar universe
only via the gravitational interactions where the ultra-lightness of the
sterile neutrino follows from the same physics that explains the near
masslessness of the familiar neutrinos.}
\section{Introduction}
There are several different observations involving neutrinos which receive a
plausible and satisfactory explanation if the neutrinos are massive.
  First is the well-known solar neutrino 
deficit\cite{solar}, observed by four different experiments\cite{expt}.  
Second is the deficit of muon neutrinos relative to electron neutrinos 
produced in the atmosphere, as measured by three experiments\cite{atmos}.  
Third is the reported evidence for $\bar{\nu}_{\mu}$ to $\bar{\nu}_e$ 
oscillation from the Los Alamos Liquid Scintillation Neutrino Detector 
(LSND) experiment\cite{LSND}. Finally, there is the likely need for a neutrino
component of the dark matter of the universe to understand the structure and
density on all distance scales\cite{HDM}. Since the highly successful
standard model of particle physics predicts zero mass for all the neutrinos, 
confirmation of any one of the above observations by ongoing and 
future experiments will already be a major step towards decoding the
nature of new physics beyond the standard model. If however all the
above findings are substantiated in future, one can reasonably expect 
the nature of this new physics to fall into only very few categories.
In this talk, I will assume the validity of all the above findings 
(although it is clear that they must be considered tentative until
further confirmation by ongoing and future experiments) and
argue first that it severely restricts the neutrino mass texture and in
particular requires the existence of a new ultra-light sterile neutrino.
I will then outline the two see-saw formulae for understanding the
small neutrino masses, and discuss their implementation in $SO(10)$
models. I will then argue that the simplest scenario which explains
the lightness of the sterile neutrino in a natural manner is one that
involves a shadow universe which has identical particle and force content
as the familiar universe but with the weak scale being somewhat higher.

\subsection{Solar Neutrino Deficit}\label{subsec:SNP}
For massive neutrinos which can oscillate from one species to another, the
solar electron neutrino observations\cite{expt} 
can be understood if the neutrino mass
differences and mixing angles fall into one of the following ranges\cite{solar},
where the Mikheyev-Smirnov-Wolfenstein (MSW) mechanism is included:
\bea
  {\rm a)}&{\rm Small-angle\ MSW,\ }\Delta m^2_{ei}\sim 5\times10^{-6}
- 10^{-5}{\rm eV}^2,
         \ \sin^22\theta_{ei}\sim7\times10^{-3},&\cr
  {\rm b)}&{\rm Large-angle\ MSW,\ }\Delta m^2_{ei}\sim9\times10^{-6}{\rm eV}^2,
         \ \sin^22\theta_{ei}\sim0.6,&(1)\cr
\eea
If the solar neutrinos oscillate into sterile neutrinos, the MSW effect
is different from the $\nu_e$ to $\nu_{\mu}$ case 
and the large angle solution is no more allowed. The above results are 
based on the approximation that only two of the neutrino species are
involved in the oscillation. 

\subsection{Atmospheric Neutrino Deficit}\label{subsec:atmos}
The second set of experiments indicating non-zero neutrino masses and mixings
has to do with atmospheric $\nu_\mu$'s and $\nu_e$'s arising from the decays of
$\pi$'s and $K$'s and the subsequent decays of secondary muons produced in the
final states of the $\pi$ and $K$ decays.  In the underground experiments the
$\nu_\mu$ and ${\bar\nu}_\mu$ produce muons and the $\nu_e$ and ${\bar\nu}_e$
lead to $e^\pm$.  Observations of $\mu^\pm$ and $e^\pm$ indicate a far lower
value for $\nu_\mu$ and ${\bar\nu}_\mu$ than suggested by na\"\i{}ve counting
arguments which imply that $N(\nu_\mu+{\bar\nu}_\mu)=2N(\nu_e+{\bar\nu}_e)$
\cite{atmos}.
If one assumes that the oscillation of $\nu_{\mu}$ to $\nu_{\tau}$ 
provides an explanation of these results, then to fits to both the
sub-GeV and multi-GeV data require that\cite{fukuda}
\bea
\Delta m^2_{\mu \tau}\approx0.025{\ \rm to\ 0.005\ eV}^2,\ 
sin^22\theta_{\mu \tau}\approx .6~~to~~1.
\eea
A recent reanalysis of the data\cite{yasuda} 
seems to imply that the data allows an upper
limit on the $\Delta m^2$ upto $.25$ eV$^2$ at 90\% confidence level.

\subsection{Results from the LSND experiment}\label{subsec:lsnd}

Recently, the LSND collaboration has published the results of their search for
$\bar{\nu}_{\mu}$ to $\bar{\nu}_e$ oscillation using the liquid scintillation
detector at Los Alamos. Combining their results which indicate a positive
result with the negative results by the E776 group and the 
Bugey reactor data, one can conclude that a mass difference
squared between the $\nu_e$ and the $\nu_{\mu}$ lies between
\bea
0.27~eV^2\leq \Delta m^2 \leq 2.3~eV^2
\eea
with points at 6 and 10 eV$^2$ also perhaps allowed.
 
\subsection{Hot Dark Matter}\label{subsec:hdm}
There is increasing evidence that more than 90\% of the mass in the universe
must be detectable so far only by its gravitational effects.  This dark matter
is likely to be a mix of $\sim20$\% of particles which were relativistic at the
time of freeze-out from equilibrium in the early universe (hot dark matter) and
$\sim70$\% of particles which were non-relativistic (cold dark matter).  Such a
mixture gives the best fit 
of any available model to the structure and
density of the universe on all distance scales, such as the anisotropy of the
microwave background, galaxy-galaxy angular correlations, velocity fields on
large and small scales, correlations of galaxy clusters, etc.  A very plausible
candidate for hot dark matter is one or more species of neutrinos with total
mass of $m_{\nu_H}=93h^2F_H\Omega=4.8$ eV, if $h=0.5$ (the Hubble constant
in units of 100 km$\cdot$s$^{-1}\cdot$Mpc$^{-1}$), $F_H=0.3$ (the fraction
of dark matter which is hot), and $\Omega=1$ (the ratio of density of the
universe to closure density).  

It is usually assumed that the $\nu_\tau$ would supply the hot dark matter.
However,
if the atmospheric $\nu_\mu$ deficit is due to $\nu_\mu\to\nu_\tau$, the
$\nu_\tau$ alone cannot be the hot dark matter, since the $\nu_\mu$ and
$\nu_\tau$ need to be closer to each other in mass. 
It is interesting that instead
of a single $\sim 4.8$ eV neutrino, sharing that $\sim 4.8$ eV 
between two or among
three neutrino species provides a better fit to the universe structure and
particularly a better understanding of the variation of matter density with
distance scale\cite{HDM}.

It is worth noting that an equally popular picture adopts 
the hypothesis
that there is a large cosmological constant ($\Omega_{\Lambda}=.8$ or so)
in a low density baryon plus CDM universe to make up  $\Omega=1$\cite{stein}. 
This has been inspired by reported large values ($h_0=.7-.8$ or so) 
of the Hubble parameter from several observations\cite{freedman} which
have hard time fitting the age of the universe (e.g. from globular
clusters) with $\Omega=1$ without a cosmological constant. 
There are however other observations that give a lower value for $h_0$
($h_0\simeq .5$). The final verdict on the dark matter picture of
the universe will therefore have to wait. It is nevertheless heartening
that there is a compelling case for a neutrino mass in the eV range
from structure formation in the universe.

In understanding the detailed implications of these data for physics
beyond the standard model, one must also take into account other constraints
on neutrinos, from nucleosynthesis, the
Heidelberg-Moscow\cite{klap} $\beta\beta_{0\nu}$ 
experiment searching for the Majorana mass
of the neutrino using enriched $^{76}$Ge and 
the synthesis of heavy elements supposedly by the rapid neutron capture
process (the so-called {\it r-process}) around supernovae\cite{fuller}.

\subsection{Other costraints:}\label{subsec:other}

\noindent (i) While the $Z^0$ width limits 
the number of weakly interacting neutrino species
to three, the nucleosynthesis limit\cite{sarkar}
of about 3.3 on the number of light
neutrinos is more useful here, since it is independent of the neutrino
interactions.  Invoking a fourth neutrino, $\nu_s$, which is sterile, meaning
it does not have the usual weak interaction, must be done with parameters such
that it will not lead to overproduction of light elements in the early
universe.  For example, the atmospheric $\nu_\mu$ problem cannot be explained
by $\nu_\mu\to\nu_s$, since $\sin^22\theta_{\mu s}\approx0.5$ is too large for
the $\Delta m^2_{\mu s}$ involved, and that $\nu_s$ would have been brought
into equilibrium in the early universe .  On the other hand, the solar
$\nu_e$ problem can be explained by $\nu_e\to\nu_s$ for either the small-angle
MSW or the vacuum oscillation solutions, but not for the less favored
large-angle MSW solution.
                                                                       
\noindent(ii) The Heidelberg-Moscow $^{76}$Ge 
experiment\cite{klap} has been conducting
a high precision search for neutrinoless double beta decay for the past
several years and have at present set the most stringent upper limits
on the effective Majorana mass of the neutrino: $<m_{\nu}>\leq .56$ eV.

\noindent(iii) It has been pointed out that in minimal model with three
massive neutrinos, supernova r-processes provide a very stringent constraint
on the neutrino mixings for eV mass range or higher. The origin of this
constraint can be understood as follows. Inside the supernova, the MSW
phenomenon enhances the conversion of the muon neutrinos (which have
higher energy) to electron neutrinos if the mass difference square
$\Delta m^2\geq 4~(eV)^2$ while leaving the $\bar{\nu}_{\mu}$'s
unaffected. The newly born high energy $\nu_e$'s deplete the neutron
content of the supernova environment via the reaction $\nu_e + n\to
e^-+p$. This reduction of the neutron content slows down the r-process
making it difficult to understand the heavy element abundance of the 
present universe. This result crucially hinges on the assumption that
$m_{\nu_{\mu}}\geq m_{\nu_e}$ and that there are neutrinos that $\nu_{\mu}$
mixes with. In fact in the presence of sterile neutrinos, its mixing
with $\nu_{\mu}$ can lead to MSW enhancement of $\nu_{\mu}$ to $\nu_s$
conversion deeper in the supernova providing a way out of this 
constraint\cite{pelto}. 

\section{Neutrino mass textures consistent with data}

In discussing the neutrino mass textures in this section,
we will assume that all the neutrinos are Majorana particles, since
it is easier to understand the smallness of Majorana masses of neutrinos
within the framework of grand unified theories.
Before going to a detailed discussion of the allowed mass matrices, let
us note two generic requirements for the allowed mass matrices dictated
by the data:{\it (i) at least two neutrinos must be degenerate in mass;
and (ii) there is a very compelling case for the existence of a sterile 
neutrino in the present data.} 

\subsection{Are neutrinos degenerate?}\label{subsec:deg}
If only two of the above hints (either
solar and atmospheric data or solar and HDM) are taken seriously, then
one can maintain a hierarchical picture for neutrino masses i.e.
$m_{\nu_e}\ll m_{\nu_{\mu}} \ll m_{\nu_{\tau}}$. Such a pattern emerges
very naturally in one class of the see-saw models (see below). However,
if we take any three of the above four hints for neutrino masses, then
we must have at least two neutrinos degenerate\cite{caldwell}. To see
the case for a sterile neutrino, let us first note that
it is not easy to write down a neutrino mass matrix within the three
generation picture that can accomodate all the above observations as well
as constraints. 
The main obstacle comes from the conflict between the 
LSND data and the MSW resolution of the solar neutrinodata.
The first requires that
the $\Delta m^2_{\nu_e-\nu_{\mu}}$ is in the eV range which is much larger than
the mass difference required to explain the solar neutrino data..
 If we ignore the LSND
data, the solar, atmospheric data and the HDM neutrino can be accomodated
in a three neutrino scenario
by assuming that all three neutrinos are degenerate in mass.\cite{caldwell}.
One very marginal possibility\cite{fuller2} has been advocated recently
using a variant of this, 
to accomodate the LSND results in this picture provided the LSND $\Delta m^2$
is chosen to be around $.3$ eV$^2$. First point to note is that since
solar neutrino puzzle requires that $\Delta m^2_{e-\mu}\simeq 10^{-5}$
eV$^2$, to understand the LSND results in this scenario, one must use 
the complete three neutrino oscillation keeping all mixing angles\cite{three}.
 This requires first that $\nu_{e}-\nu_{\tau}$ mixing angle is not too small.
Secondly, we must have $\Delta^2_{\mu-\tau}$ be $\approx .3~eV^2$.
Thus the oscillation frequency is determined by $\nu_e$-$\nu_{\tau}$
mass difference. The main problem
for this scenario comes from the atmospheric neutrino data,
since the original analysis of the Kamiokande sub-GeV and 
the multi-GeV data by the Kamiokande group excludes $\Delta m^2\geq .1~eV^2$
at 90\% confidence level (c.l.). As mentioned earlier, a subsequent analysis
\cite{yasuda} extends this range further excuding only
the $\Delta m^2\geq 0.25~eV^2$ at 90\% confidence level while allowing it
at 95\% c.l. level. While at its face a value of $m_{\nu_e}\simeq 1.6$ eV
may appear to be in conflict with the neutrinoless double beta decay limit
\cite{klap}, one can hide under the uncertainties of nuclear matrix element
calculations which typically could be as much as a factor of 2-3. 
As the precision in $\beta\beta_{0\nu}$ search improves further (say
to the level of $0.1$ eV), nuclear matrix element uncertainties cannot
come to the rescue and this mass texture will then be ruled out. One can 
write the neutrino Majorana mass matrix for this case as follows:
\bea
M=\pmatrix{m+\delta_1s^2_1&-\delta_1c_1c_2s_1&-\delta_1c_1s_1s_2\cr
             -\delta_1c_1c_2s_1&m+\delta_1c^2_1c^2_2+\delta_2s^2_2&
             \delta_1c^2_1c_2s_2-\delta_2c_2s_2\cr
             -\delta_1c_1s_1s_2&\delta_1c^2_1s_2c_2-\delta_2c_2s_2&
             m+\delta_1c^2_1s^2_2+\delta_2c^2_2\cr},
\eea
where $c_i=\cos\theta_i$ and $s_i=\sin\theta_i$, $m=1.6$ eV;
$\delta_1\simeq 1.5\times 10^{-5}$ eV;
 $\delta_2\simeq.1$ eV; $s_1\simeq0.05$; and
$s_2\simeq0.4$ for the small-angle MSW solution.

\subsection{The need for a sterile neutrino}\label{subsec:sterile}

\noindent We thus see that if the above scenario is ruled out,
for instance by the tightening of the double beta decay limit on
the Majorana mass of $\nu_e$ or by the atmospheric neutrino data, then the
only way to understand all neutrino results 
will be to assume the existence of an
additional neutrino species which in view of the LEP data must not couple
(or couple extremely weakly) to the Z-boson.We will call this the
sterile neutrino. The picture then would be as
follows\cite{caldwell,valle}: the solar neutrino puzzle is explained by the
$\nu_e-\nu_s$ oscillation; atmospheric neutrino data would be explained by
the $\nu_{\mu}-\nu_{\tau}$ oscillation. The LSND data would set the
overall scale for the masses of $\nu_{\mu}$ and $\nu_{\tau}$ (which are 
nearly degenerate)  and if this
scale is around 2 to 3 eV  as is allowed by the data\cite{LSND}, then
the $\nu_{\mu,\tau}$ would constitute the hot dark matter of the universe.
The mass matrix in this case would be
 in the basis ($\nu_s$, $\nu_e$, $\nu_\mu$, $\nu_\tau$),
\begin{eqnarray}
M=\pmatrix{\mu_1&\mu_3&\epsilon_{11}&\epsilon_{12}\cr
             \mu_3&\mu_2&\epsilon_{21}&\epsilon_{22}\cr
             \epsilon_{11}&\epsilon_{21}&m&\delta/2\cr
             \epsilon_{12}&\epsilon_{22}&\delta/2&m+\delta\cr}.
\end{eqnarray}
For simplicity, we set the $\epsilon_{12}=\epsilon_{22}=0$ and $\mu_2\ll \mu_1
\simeq 10^{-3}~eV$.
The $\epsilon_{11}$ term is responsible for the $\nu_e-\nu_{\mu}$
oscillation that can explain the LSND data. The apparent problem
for such a scenario comes from the supernova r-process nucleosynthesis.
But it has been argued\cite{pelto} that in such a scenario, the $\nu_{\mu}$
can oscillate into the $\nu_s$ at a smaller protoneutron star radius before
it reaches the radius where $\nu_{\mu}$ to $\nu_e$ MSW transition occurs.
This may enable one to evade the r-process bound for $\Delta m^2_{e-\mu}\geq
4~eV^2$. Clearly the crucial test of the sterile neutrino scenario will
come when SNO collaboration obtains their results for neutral current
scattering of solar neutrinos. One would expect that $\Phi_{CC}=\Phi_{NC}$
if the $\nu_e$ oscillation to $\nu_s$ is responsible for the solar neutrino
deficit. There should be no signal in $\beta\beta_{0\nu}$ search. Precision
measurement of the energy distribution in 
charged current scattering of solar neutrinos at Super-Kamiokande can also
shed light on this issue.

Before proceeding to the discussion of the theoretical implications
of the mass textures outlined above, we want to note that if the atmospheric
neutrino data is excluded but LSND, HDM and solar neutrino constraints 
are kept, a theoretical explanation for them can be found also with
an inverted mass texture\cite{silk} for neutrinos where the
$m_{\nu_e}\simeq m_{\nu_{\tau}}\simeq 2.4~eV\gg m_{\nu_{\mu}}$ and
which does not invoke the sterile neutrino. This texture is consistent
with the supernova r-process constraints and uses the $\nu_e \to \nu_{\tau}$
large angle MSW solution to explain the solar neutrino data. This could
therefore be tested once Super Kamiokande results for the neutrino
energy spectrum as well as the data on day-night variation is in.  

\section{Implications for higher unification and two types of
see-saw mechanism}

In this section we address the question of what implications the 
small nonzero neutrino masses
and in particular any of the scenarios discussed above have for the nature
for the nature of new physics beyond the standard model. To start with
let us remind the reader that in the standard model the presence of an
exact global B-L symmetry combined with the absence of the right handed
neutrino leads to zero mass for all neutrinos. The simplest way to generate
a nonzero neutrino mass is therefore to add three right handed neutrinos
$N_i$, one per generation. It is easy to see that as soon as the $N_i$ are
included, the maximal anomaly-free gaugeable symmetry becomes
$SU(2)_L\times SU(2)_R\times U(1)_{B-L}\times SU(3)_c$ which can eventually
lead to an $SO(10)$ grand unification of fermions. As has been shown
during the past decade and half, this class of models provide the most
natural framework for describing the neutrino masses\cite{gellmann,goran}.
What happens in these models is that as the $SU(2)_R\times U(1)_{B-L}$
gauge group breaks down to $U(1)_Y$, not only the right-handed gauge bosons
but also the right handed neutrinos of all three generations acquire a mass
$fv_R$ proportional to the $B-L$ breaking scale $v_R\gg v_W$, where $v_W$ is the
electroweak symmetry breaking scale of the standard model. At this stage,
the left handed neutrinos are massless. At the scale $v_W$, the Dirac 
mass for the neutrino that connects the left and the right handed neutrino
is generated with a value given by $hv_W$ which is expected to be of the
order of the masses of the charged fermions $m_f$ which also arise at that scale.
This leads to the see-saw matrix for the neutrinos\cite{yana,gellmann,goran},
\bea
M=\left(\begin{array}{cc}
0 & m_f\\
m_f & M_N
\end{array} \right)
\eea
The diagonalization matrix  leads to the generic formula for neutrino masses
\bea 
m_{\nu_i}\simeq {{(m_f)^2}\over{M_{N_i}}}
\eea
This formula has two interesting implications: (i) the first is that the
neutrinos, which are now necessarily Majorana fermions have masses which
are suppressed compared to the masses of the charged fermions of the
corresponding genearation and (ii) the neutrino masses show a 
generationwise hierarchical
pattern linked to the square of the masses of the charged fermions of the
corresponding generation (i.e. $m_{\nu_e}\ll m_{\nu_{\mu}}\ll m_{\nu_{\tau}}$)
We will call this the type I see-saw formula.

It was pointed out in \cite{goran} that when the spontaneous symmetry
breaking of the left-right model (or its $SO(10)$ grand unified version)
is carefully analyzed, one actually gets a modified neutrino mass matrix
given by
\bea
\left(\begin{array}{cc}
 {{\lambda v^2_W}\over{v_R}} & m_f \\
m_f & M_N
\end{array}\right)
\eea
Diagonalization of the above mass matrix leads to what I call the type II
see-saw formula for the neutrino masses:
\bea
m_{\nu_i}\simeq {{\lambda v^2_W}\over{v_R}}-{{m^2_f}\over{M_{N_i}}}
\eea      
Note that the first term in the above formula is practically generation
independent. Therefore, the neutrino mass pattern in this case is not
hierarchical and could lead to a nearly degenerate spectrum as has been
advocated in the previous section.

There are conditions under which the type II see-saw formula reduces to
a type I see-saw formula\cite{chang}: for instance when the discrete
parity symmetry of the left-right or $SO(10)$ models is broken at a
scale higher than the $SU(2)_R\times U(1)_{B-L}$ gauge symmetry, then
the first term in the type II see-saw formula is replaced by 
$\lambda v_W v^2_R/M^2_P$ ($M_P$ being the scale of discrete parity
breaking), which can clearly be arranged to yield the hierarchical
mass pattern. Another class of models where the type I see-saw can
emerge are some supersymmetric models with restricted Higgs representations.

A very interesting point worth emphasizing here is that if we look at
the typical masses needed to solve the solar as well as the atmospheric
neutrino puzzles and use the see-saw formula to find the scale of $B-L$
symmetry breaking, we find that $v_R\approx 10^{12}-10^{13}$ GeV.
It may be more than a mere coincidence that the B-L breaking scale of
$v_R\sim10^{13}$ GeV emerges naturally from constraints of
$\sin^22\theta_W$ and $\alpha_s$ in non-supersymmetric SO(10) grandunified
theories\cite{chang1}, as well as supersymmetric SO(10)\cite{lee3} theories.
In the least it enhances the reason for an SO(10) scenario.
It is however possible to construct TeV scale right handed neutrino 
scenarios\cite{babu1} where the suppression of the neutrino mass originates
from the fact that the Dirac masses are radiatively induced.
To summarize this section, it is reasonable to conclude that evidence for
a small neutrino mass would indicate the existence of a local $B-L$ 
symmetry in nature and perhaps even a left-right symmetry, which will
be a major new dimension to our understanding of particle physics.
Secondly, the generic class of grand unified models where
the see-saw mechanism (both type I and type II) is naturally implemented
are based on the $SO(10)$ GUT group with the type I see-saw leading to
a hierarchical pattern for neutrino masses whereas the type II leads to
a near degenerate pattern. In the next section, we explore
whether definite predictions can be made for the neutrino masses and mixing
angles in this class of models.

\section{Predictions from minimal SO(10) grand unification models }
If there are only three light neutrinos, it is both economical and elegant
to work within simple $SO(10)$ grand unified models. 
While simple electroweak gauge theories without additional symmetries
do not have the capability to predict fermion masses, the assumption of
 grand unification improves this record somewhat (e.g. the
prediction of b-quark mass in $SU(5)$). In the minimal $SO(10)$ models,
the neutrino Dirac mass and the up-quark mass matrices become equal
since they both arise from the Yukawa couplings of the fermions
(which belong to the {\bf 16} dim representations) to the Higgs boson
in the {\bf 10}-dim representation, thereby reducing the number of
free parameters. This raises
the possibility for a prediction of the neutrino masses in these models. 
The problem however is that the Majorana mass of the $N_i$ arises from the
couplings of the fermions to the $\bar{\bf 126}$ dimensional Higgs bosons.
Since these couplings are arbitrary, in general no specific predictions
can be made. It was however pointed out by Babu and this author\cite{babu}
that in the minimal $SO(10)$ models, the standard model doublets arise
from an admixture of the $SU(2)_L$ doublets in {\bf 10} and $\bar{\bf 126}$
dimensional Higgs bosons. Therefore, the $\bar{\bf 126}$ Yukawa couplings
(as well as those to {\bf 10}) get predicted in terms of the quark, lepton
masses and their mixings. This model (which is a realization of the
type I see-saw mechanism) therefore leads to numerical predictions
for the neutrino masses and mixings. The reader is refered to the original
papers for the detailed predictions for the non-supersymmetric\cite{babu,lav}
as well as supersymmetric versions\cite{lee} of the model. 
There are actually six solutions depending on the relative signs of the
various quark masses. Here we simply want
to note that there are predictions in both versions that can accomodate the
small angle MSW solution to the solar neutrino puzzle but not the atmospheric
nor the LSND nor the HDM neutrino.   

    There is another class of $SO(10)$ 
models\cite{achiman} where additional symmetries
are imposed to fix the heavy Majorana mass matrix for the right handed 
neutrinos and different popular quark mass textures are used for the
neutrino Dirac masses. They also implement the type I see-saw formula
and give generic predictions that can accomodate only the small angle 
MSW solutions to the solar neutrino puzzle.

Finally a different class $SO(10)$ models were studied\cite{lee1} where
the type II see-saw mechanism was implemented. Using an additional
$S_4$-permutation symmetry on the fermions and the Higgs bosons, it
was possible to obtain a realization of the degenerate neutrino mass
mixing angle predictions that can solve both the solar as well as the
atmospheric neutrino problem.  

\section{Beyond grand unification: Into the shadow universe}
Once we admit the possibility of light sterile neutrinos, one needs to go
beyond simple grand unified models to understand why the sterile neutrino is
so light. The reason for this is that the sterile neutrino by definition is an
$SU(2)_L\times U(1)_Y$ singlet and therefore is allowed to have an arbitrary
mass unless there are some compelling new symmetries that keep it massless.
Attempts have been made using additional $U(1)$'s and supersymmetry\cite{chun}
etc to achieve this goal.
But it is perhaps fair to say that there are no compelling motivations for
such symmetries. To circumvent such arguments, it was proposed in \cite{bere}
to make the conjecture that there is an exact duplication of the standard model
in both the gauge as well as the fermion content i.e. an extra $G'_{standard}$
with $Q', {u^c}', {d^c}', L', {e^c}'$ etc. (this adds a new sector to
the world of elementary particles, which will be called
the shadow sector). It is then clear that we have three
extra neutrinos which do not interact with the Z-boson. We further assume
that the only interactions that connect the known and the shadow sector
is the gravitational interaction.

Within this framework, it is easy to understand that the shadow neutrinos
(which will be the sterile neutrinos) are massless in the renormalizable
theoryt for exactly the same
reason that the ordinary neutrinos are  (i.e. the existence of
a $B'-L'$ symmetry in the shadow standard model sector). We may assume that
there is a "shadow" see-saw mechanism which operates exactly the 
same way to give tiny masses to the shadow neutrinos. The next question is how
do the shadow (or sterile) neutrinos acquire small masses and
 mix with the known neutrinos ?
Here we use the existing lore that all global symmetries
are broken by Planck scale effects. It was already pointed out\cite{ellis}
that one can write Planck scale induced operators such as $LH_u LH_u/M_P$,
$LH_uL'H'_u/M_P$ and $L'H'_uL'H'_u/M_P$ which violate both $B-L$ as well as
$B'-L'$ symmetries and after electrweak symmetry breaking in both the
sectors lead to $\nu-\nu'$ mixing. If we now make the additional assumption
that $v'_{W}\simeq 30 v_W$, the resulting $\nu_e-\nu'_e$ mass matrix
gives a solution to the solar neutrino puzzle with small mixing angles.
When this idea is combined with the postulate that there exists an
$L_e+L_{\mu}-L_{\tau}$ symmetry (instead of the overall $B-L$ symmetry)
that is broken by Planck scale effects, we come up with a neutrino
mass matrix that explains all neutrino puzzles using the four neutrino
mass texture noted in Ref.\cite{caldwell}.

The next interesting feature of these models is that if the $m_{\nu_{\mu}}
\simeq m_{\nu_{\tau}}\simeq 2$ eV or so, then the $m_{\nu'_{\mu}}\simeq
m_{\nu'_{tau}}\simeq 2$ keV. Thus $\nu'_{\mu,\tau}$ can qualify as warm
(or cool) dark matter of the universe, a possibility which does not appear to
have been ruled out present cosmological observations. Such models have
many interesting implications for cosmology\cite{teplitz}, which we will
not go into here.                                                             

\section{Conclusions}
The solar, atmospheric and LSND neutrino data, 
along with a need for some hot dark
matter, if all are due to neutrino mass have two very profound implications:
(i) at least two neutrinos must be degenerate in mass, a feature nor shared by
charged fermions and not expected in the minimal SO(10) type models;
(ii) there is a very good possibility that there is need for a sterile
neutrino. In this talk, I have considered the various implications of
these conclusions for physics beyond the standard model, such as the
simple $SO(10)$ scenarios and conclude that one needs to go beyond such 
simple models if all the present indications neutrino mass are correct.
I then outline the recent suggestion of Z. Berezhiani and this author
that a scenario with a shadow universe with identical gauge and fermion
structure (but with an asymmetric weak scale) can explain all the neutrino
puzzles without the need for any other ingredients.
                                                   
\vskip 5mm

\noindent{\bf Acknowledgement}

This work is supported by a grant from the National Science Foundation.
I also want to thank the organizers of the Neutrino'96 conference for
financial support.

\end{document}